\documentclass[12pt]{article}

\usepackage{scicite}

\usepackage{times}

\usepackage{ amssymb }

\usepackage{graphicx}

\newenvironment{sciabstract}{%
\begin{quote} \bf}
{\end{quote}}

\newcounter{lastnote}

\title{Dynamics and Control of Bubble-Propelled Microrobots}

\author
{David P. Rivas,$^{1}$ Max Sokolich,$^{1}$ Harrison Muller, $^{1}$ Sambeeta Das, $^{1\ast}$\\
\\
\normalsize{$^{1}$Department of Mechanical Engineering, University of Delaware,}\\
\normalsize{130 Academy Street, Newark DE, USA}\\
\\
\normalsize{$^\ast$To whom correspondence should be addressed; E-mail:  samdas@udel.edu.}
}

\date{}

\begin{document} 


\baselineskip24pt


\maketitle


\begin{sciabstract}
Having the advantage of being relatively fast and powerful, as well as readily fabricated, spherical bubble-propelled microrobots are particularly well suited for applications such as cargo delivery, micromanipulation, and biological or environmental remediation. However, there have been limited examples of control and manipulation with these microrobots and few studies on their dynamics. Here we investigate the bubble formation and dynamics of both hemispherically coated Janus microrobots as well as GLAD "patchy" microrobots which not only provide for an interesting comparison, but also exhibit useful properties in their own right. Specifically, we find that the patchy microrobots have a greater tendency to produce small bubbles which is also associated with smoother motion. These properties are beneficial for using these microrobots for precise micro-manipulation, for example. We demonstrate manipulation and assemble of passive spheres on a substrate as well as at an air-liquid interface. We also characterize the propulsion and bubble formation of both types of microrobots and find that previously proposed theories are insufficient to adequately describe their motion and bubble bursting mechanism. 
\end{sciabstract}

% In setting up this template for *Science* papers, we've used both
% the \section* command and the \paragraph* command for topical
% divisions.  Which you use will of course depend on the type of paper
% you're writing.  Review Articles tend to have displayed headings, for
% which \section* is more appropriate; Research Articles, when they have
% formal topical divisions at all, tend to signal them with bold text
% that runs into the paragraph, for which \paragraph* is the right
% choice.  Either way, use the asterisk (*) modifier, as shown, to

\section*{Introduction}

Microrobots show promise in biomedical and engineering applications such as drug delivery, environmental remediation, enhanced biofilm removal, sensing, and micro-manipulation \cite{Nelson2010,PengSmall2020,Jang2019,EskandarRSC2017,Villa2020,WangACSAMI2019,SittiLabChip2017,SanchezElectroanalysis2017,ZareiSmall2018,RenSciAdv2019}. A commonly used type of microrobot is one that is catalytically driven and operates by the chemical decomposition of a fuel source present in the medium. Two common modes of propulsion rely on self-phoretic or bubble generation mechanisms \cite{Moran2017,Wang2011,Liu2017,WangLangmuir2018,ManjarePRL2012}. Bubble-propelled microrobots have the advantage of being relatively fast and powerful compared to their self-phoretic counterparts, features that are advantageous in many of the aforementioned applications. Specifically, these properties are particularly useful in tasks such as transport of cargo or rapid remediation or microassembly. However, there have been limited examples of precise control and manipulation using bubble-propelled microrobots \cite{SolovevAFM2010}, as well as relatively few studies on their dynamics and bubble formation/dissolution process \cite{WangLangmuir2018,ManjarePRL2012,ManjareJPC2013,WredeSmall2021}. There have also been limited studies on the temperature dependence of microrobot dynamics \cite{SanchezJACS2011}, particularly at values near room temperature. Exploring the temperature dependence of their dynamics provides insight into how these changes effect bubble generation and motion in realistic environmental conditions. It also connects with the concept of using microrobots as sensors to reveal properties of their environment.

Motion driven by self-phoretic effects, such as self-electrophoresis or self-diffusiophoresis, derive from a concentration gradient of either ionic or neutral molecules, respectively. This gradient forms around the particle due to an anisotropic distribution of catalyst material on the microrobots surface. A typical example of this is a Janus spherical colloid in which one hemisphere of the sphere is coated in platinum. The platinum catalyzes hydrogen peroxide to form water and oxygen which results in a concentration buildup of molecules around the colloid. n bubble-driven motion, the concentration of oxygen reaches saturation in the medium, leading to a bubble nucleating and growing \cite{Wang2011}. The propulsion of the microrobot in this case is due to the growth of the bubble, the hydrodynamic flows created upon its dissolution, or other mechanisms \cite{WredeSmall2021,ManjarePRL2012,WangLangmuir2018}.  

A type of bubble-propelled micromotor that has been extensively studied is a microjet, a hollow tubular micromotor that produces bubbles inside the tube and ejects them out one end, propelling it toward the other \cite{WredeSmall2021,ManjareJPC2013}. Microjets typically have lengths of 50-100 $\mu$m, diameters of 5-10 $\mu$m, and speeds of 300-3000 $\mu$m/s \cite{Wang2011}. This type of microrobot was previously used to transport spheres and manipulate micro-sheets at the liquid-air interface, but lacked precise magnetic control \cite{SolovevAFM2010}.  One disadvantage of the microjet is its more complex fabrication process compared to spherically shaped Janus particles which can be made using simple physical vapor deposition procedures. %[other disadvantages to the microjet?]. 

For a spherical Janus catalytic particle, diffusiophoretic motion generally occurs when the diameter of the micromotor is less than about 10 $\mu$m \cite{ManjarePRL2012}. For larger diameter microrobots, bubbles begin to form at their surface due to the greater concentration of reactant products. As previously mentioned, diffusiophoretic micromotors move at much slower speeds and are significantly less powerful than bubble-propelled microrobots, making them less ideal for applications requiring manipulation of objects at the microscale, such as microassembly. For example, bubble propelled micromotors are capable of an order of magnitude larger velocity, orders of magnitude increase in generated propulsion force, and much higher efficiency \cite{ZhangMicromachines2017}. However, bubbles generated at their surface could potentially interfere and limit their precision.

Here we study the dynamics and bubble growth process of both hemispherically coated Janus as well as GLAD "patchy" spherical bubble propelled microrobots as a function of hydrogen peroxide concentration and temperature. Specifically, we study the propulsion mechanisms of the microrobots as well as their bubble size and growth frequency and compare the results to previous theories and experimental reports. We also demonstrate assembly, manipulation, and transport of passive spheres both on a substrate and at an air-liquid interface using magnetically steered bubble-propelled microrobots. Additional precision is obtained using the patchy microspheres which tend to produce smaller bubble sizes and lack the forward/backward cyclical motion present in the case of the hemispherically coated Janus particles. This makes the patchy particles well suited for precise manipulation where a smaller bubble size and smoother motion of the microrobot is beneficial. Moving at the liquid-air interface, the micromotors could also make for useful tools in applications at interfaces such as in environmental remediation or localized interfacial biofilm removal. 

\begin{figure}
  \centering
  \includegraphics[width=.75\linewidth]{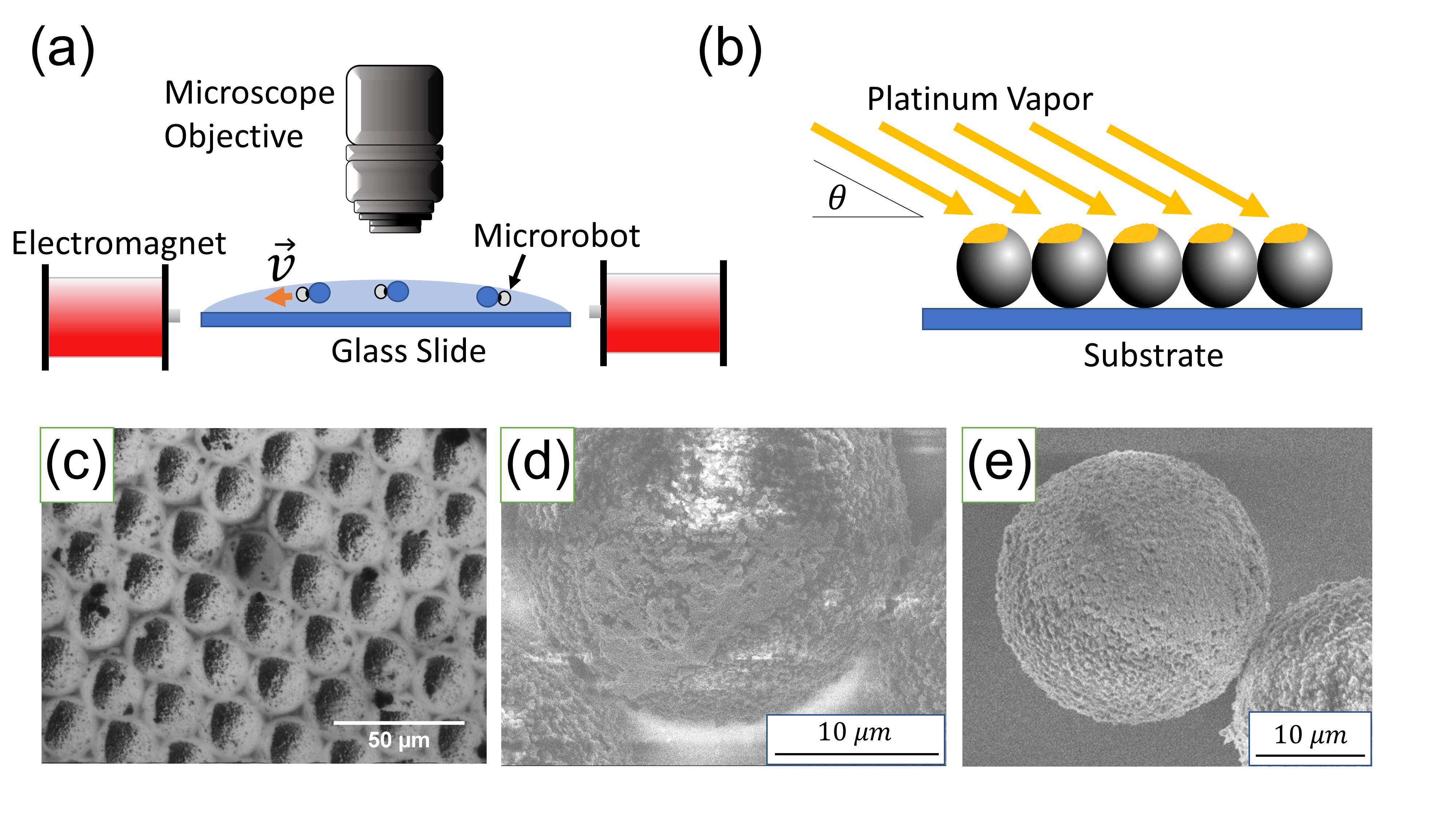}
  \caption {(a)	Experimental schematic showing the buoyant bubble propelled microrobots in hydrogen peroxide solution and two of the four electromagnets used to steer them. (b) A sketch of the glancing-angle deposition onto the densely packed spherical particles. The angle of deposition is denoted by $\theta$. (c) A fluorescent image of the spheres after the glancing-angle deposition of platinum on their surface. (d) SEM image of one of the “patchy” spherical particles. (e) SEM image of a hemispherically coated sphere.}
  \label{intro}
\end{figure}

Figure ~\ref{intro}(a) shows a schematic of the experimental setup. The microrobots were placed in a solution of hydrogen peroxide which was spread onto a hydrophilic glass slide. Electromagnets were used to steer the magnetic microrobots. Microrobots were fabricated using e-beam deposition, as depicted in Figure ~\ref{intro}(b).The spheres were first coated with iron at a deposition angle of $\theta = 90^{\circ}$ and then coated in platinum at an angle of $\theta = 10^{\circ}$ or $\theta = 90^{\circ}$, to create "patchy" microrobots and hemispherically coated microrobots, respectively.  Figure ~\ref{intro}(c) shows a fluorescence image displaying the platinum coating on the surface of the patchy microrobots. Figure ~\ref{intro}(d) and (e) display SEM images of the patchy and hemispherically coated spherical particles, respectively. The surface coverage of the patchy microrobots is approximately 1/4 of that of the hemispherically coated microrobots \cite{PawarLangmuir2008}. More details on the fabrication and experimental procedures can be found in the Materials and Methods section. %This surface coverage is similar to that of a sphere that is about half the diameter, which is near the expected threshold diameter for bubble formation. One potentially important distinction, however, is the difference in the surface curvature and the non-spherically symmetric surface coating of the patchy microrobots compared to that of a smaller spherical microrobot, which may lead to differences in the generation or dissolution of the bubbles.

We study both the patchy and hemispherically coated microrobots at hydrogen peroxide concentrations of 10, 15, and 30$\%$, as well as at temperatures of 67, 71, 76, and 80 degrees Fahrenheit. Experiments done at the various hydrogen peroxide concentrations were conducted at a temperature of approximately 75 ${^\circ}$F and the temperature controlled experiments were all conducted at a hydrogen peroxide concentration of 30$\%$. 

Microrobots were suspended in a solution of hydrogen peroxide and then viewed through a microscope on a glass slide. The platinum on the microrobots surface acts as a catalyst of the hydrogen peroxide solution, resulting in the formation of oxygen bubbles. The bubbles cause the microrobots to become buoyant and rise to the air-liquid interface, as has been reported previously with bubble-propelled microrobots \cite{WangLangmuir2014}. The microrobots generate bubbles in a cyclical manner that follows a pattern of bubble growth followed by bursting, causing the microrobots to self-propel at the air-liquid interface. 

In the following sections, we begin by describing experiments in which we used the microrobots to precisely manipulate passive spheres both at air-liquid interface and on the glass substrate. We then describe the bubble generation and dynamics of both the hemispherically coated and patchy microrobots. We then discuss possible mechanisms that determine the bubble size and frequency of bursting. This is followed by a description of additional interesting observations of microrobot behavior, and lastly we end with a discussion of our results and possible future experiments and applications.

\subsection{Manipulation and Assembly}
\begin{figure}
  \centering
  \includegraphics[width=.75\linewidth]{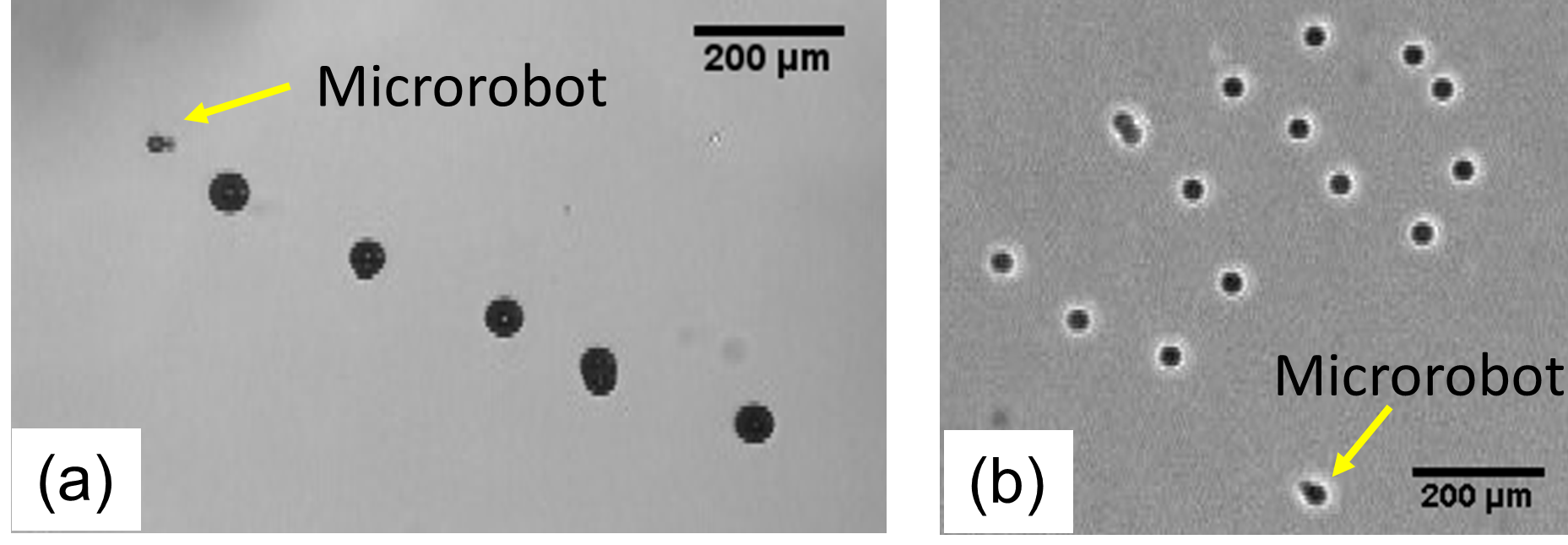}
  \caption { (a) A bubble-propelled microrobot assembled buoyant 45-85 $\mu$m diameter spheres into a linear shape at the liquid-air interface. The microrobot is magnetically steered and is coated with platinum by e-beam deposition at a glancing-angle of 10 degrees. The concentration of hydrogen peroxide was 30 $\%$ (b) Images showing manipulation of passive 24 $\mu$m diameter spheres into the letters “UD” on the glass substrate by a microrobot. The microrobot is coated in iron to make it magnetic and then with platinum at a 20 degree glancing angle. The concentration of hydrogen peroxide was 20$\%$. }
  %([videos are called 80degree_30%H2O2_TiO2HollowSphereswithbots.avi and 24um_20%H2O2_UD. Note that I have images of a better U and D but they didn’t exist at the same time, not sure if that’s better])%
  \label{manipulation}
\end{figure}

We demonstrate magnetic control of the microrobots by manipulating passive buoyant hollow 45-85 $\mu$m diameter spheres which float at the liquid-air interface. We are able to be assemble the hollow spheres into shapes by the microrobots, as shown in Vid. 6 in the SI and Fig.~\ref{manipulation}(a). The spheres are manipulated by directing the microrobots such that they collide with the spheres, or by placing them in near proximity to the spheres so that the fluid flow generated by the moving microrobots would exert a force on the spheres. We also demonstrate precise manipulation of smaller 24 $\mu$m diameter spheres on the substrate (see Vid. 7 in the SI and Fig.~\ref{manipulation}(b)). Manipulation at the substrate was made possible by using a thin layer of liquid on the glass slide, so that the buoyant microrobots could physically interact with the spheres on the substrate. We note that, to the best of our knowledge, such precise control and micromanipulation using bubble propelled microrobots has not been demonstrated previously, particularly at the air-liquid interface. %Assembly of the passive spheres into other designs is shown in Vid.  in the SI.

\subsection{Microrobot Velocity and Bubble Production}
\begin{figure}
  \centering
  \includegraphics[width=.95\linewidth]{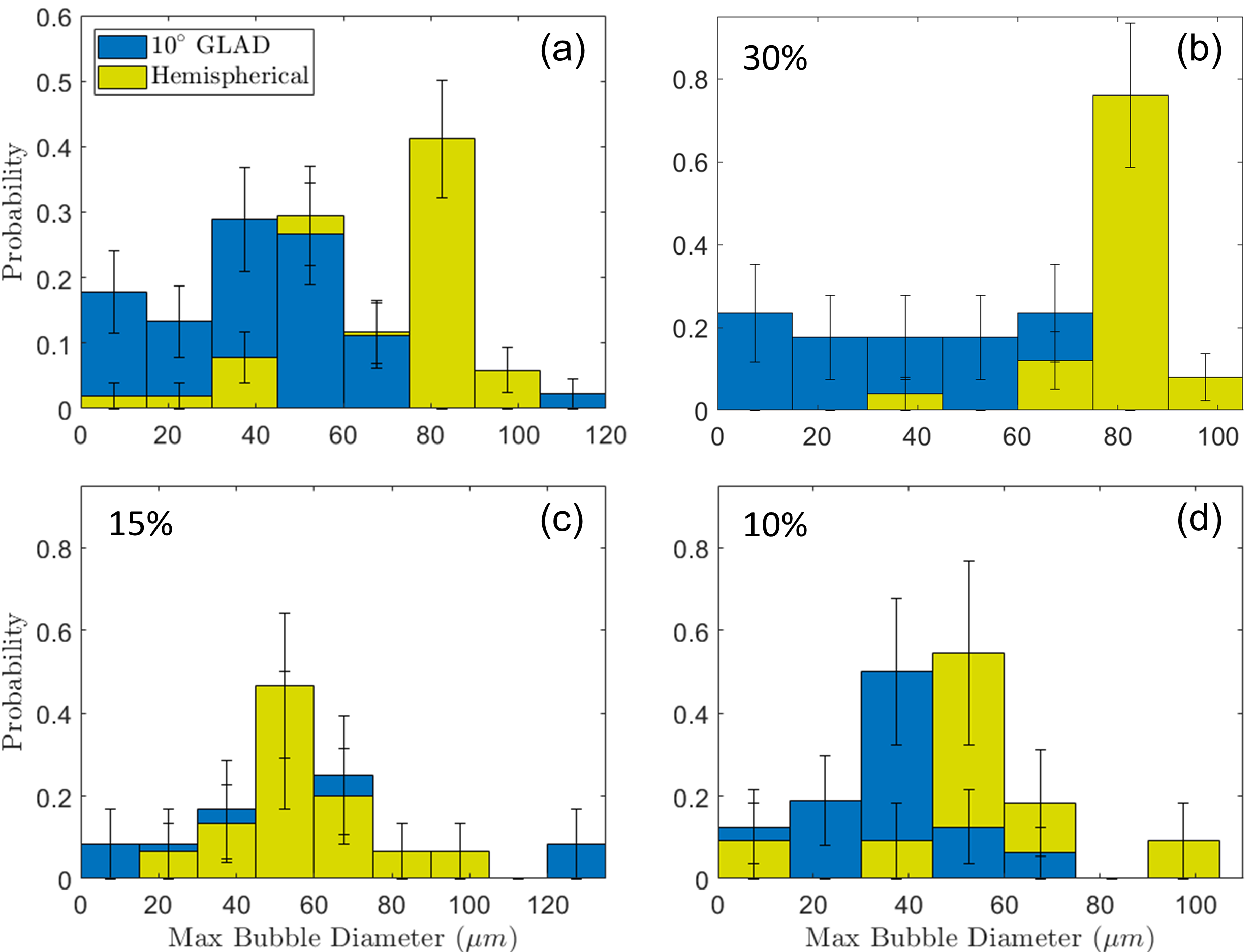}
  \caption {Probability of a micromotor producing a bubble of a given diameter for both the patchy and hemispherically coated microrobots for (a) all hydrogen peroxide concentrations, (b-d) concentrations of 10, 15, and 30$\%$, respectively. The larger bar is placed behind the smaller. The error bars correspond to square root of N uncertainties.}
  \label{bubblehisto}
\end{figure}

We find that the microrobots exhibit a wide range of bubble sizes, bubble burst frequencies, and velocities. Despite this variability, any given microrobot generally produces bubbles at a remarkably consistent rate and size. The bubbles produced by the microrobots range from less than a few microns in diameter up to approximately $D_{max} = 120 \mu m$ in diameter. In Fig.~\ref{bubblehisto} we show the probability distribution of the maximum bubble diameter for all hydrogen peroxide concentrations used. As can be seen from the figure, the patchy microrobots generally produce bubbles of smaller size than those produced by the hemispherically coated microrobots, particularly at the highest concentration of hydrogen peroxide of 30$\%$. Indeed, a subset of the patchy microrobots produce bubbles of diameter similar or smaller than the diameter of the micromotors themselves. This could be a useful feature when precise control of the placement of the microrobots is desired while avoiding disturbances created by the physical interference of larger bubbles or the hydrodynamic influence created upon its bursting. The larger variability in maximum bubble size of the patchy compared to the hemispherically coated microrobots could be due to slight differences in the shape or surface area of the platinum coating on the patchy colloids. This is partly due to the spheres not forming a perfect close-packed monolayer prior to deposition, differences in deposition orientations relative to the hexagonal structure, and polydispersity of sphere sizes. 

\begin{figure}
  \centering
  \includegraphics[width=.95\linewidth]{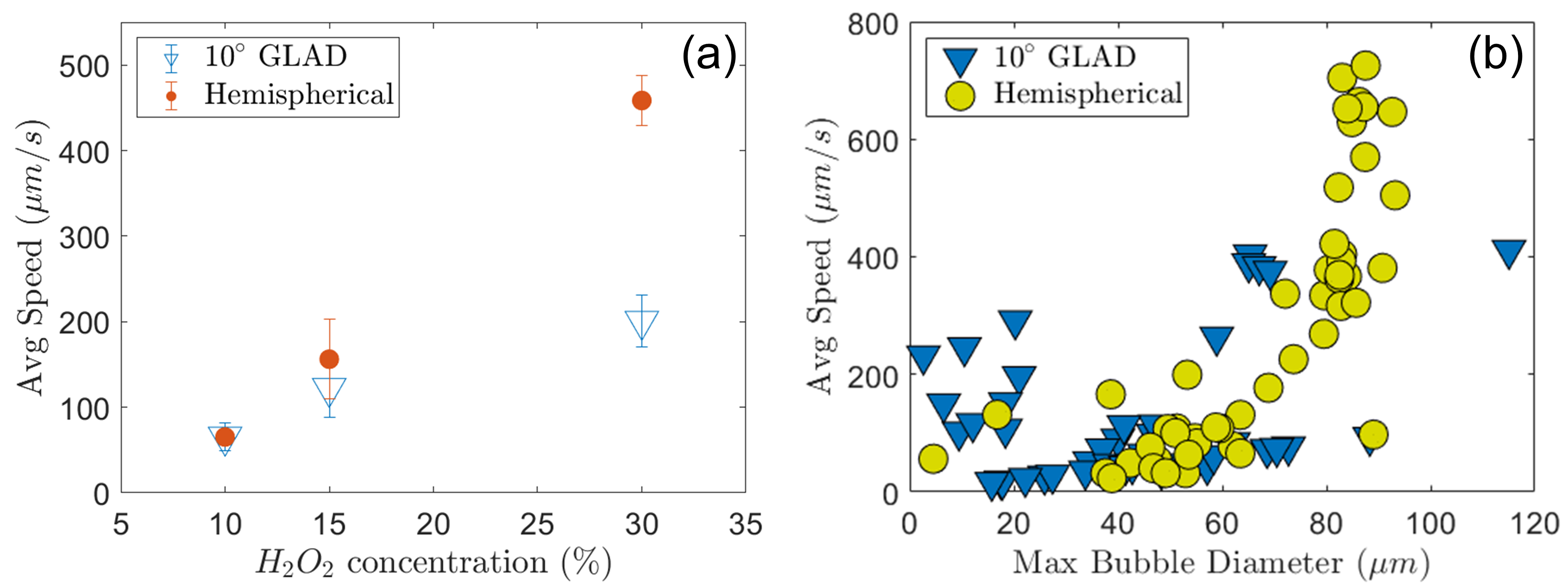}
  \caption {(a) Average speed of the patchy (blue triangles) and hemispherically coated (red circles) bubble-propelled microrobots as a function of hydrogen peroxide concentration. (b) The average speed of the microrobots versus their maximum bubble diameter, for all hydrogen peroxide concentrations.}
  \label{VelVsBubbleSz}
\end{figure}

One possible explanation for the decreased bubble size observed in the case of the patchy compared to the Janus microrobots could be the reduced platinum coating, resulting in a lower oxygen production rate. One would then expect that the Janus microrobots would have the same small bubble sizes if a sufficiently lower hydrogen peroxide concentration was used, hence also lowering the oxygen production rate. As can be seen from Fig.~\ref{bubblehisto}, lowering the hydrogen peroxide concentration does result in a reduction of the bubble size of the Janus microrobots, although the probability for the smallest bubble sizes (the two left-most bins, for example) remains lower than that of the patchy microrobots. We also find that the speed of the Janus microrobots decreases substantially as the hydrogen peroxide concentration is decreased, as shown in Fig.~\ref{VelVsBubbleSz}(a). Further reducing the hydrogen peroxide concentration to 5 $\%$ resulted in velocities that were not sufficiently larger than drift velocities to be accurately measured. We will discuss what factors might determine the bubble size and frequency for these microrobots in section \ref{bubblesz}.

We find that the velocity of the microrobots generally increases with the maximum size of the bubble produced, see Fig.~\ref{VelVsBubbleSz}(b), up to a maximum diameter of about 100 $\mu m$ at which point the speed appears to have little to no dependence on bubble size. Micrororobots that produce bubble diameters of less than about 25 $\mu m$ tend to have relatively large speeds compared to those that produce somewhat larger bubbles of 25-50 $\mu m$ in diameter. This may indicate a distinction in their bubble production/burst or propulsion mechanisms, which we will discuss later.

\subsection{Microrobot Dynamics}

\begin{figure}
  \centering
  \includegraphics[width=.85\linewidth]{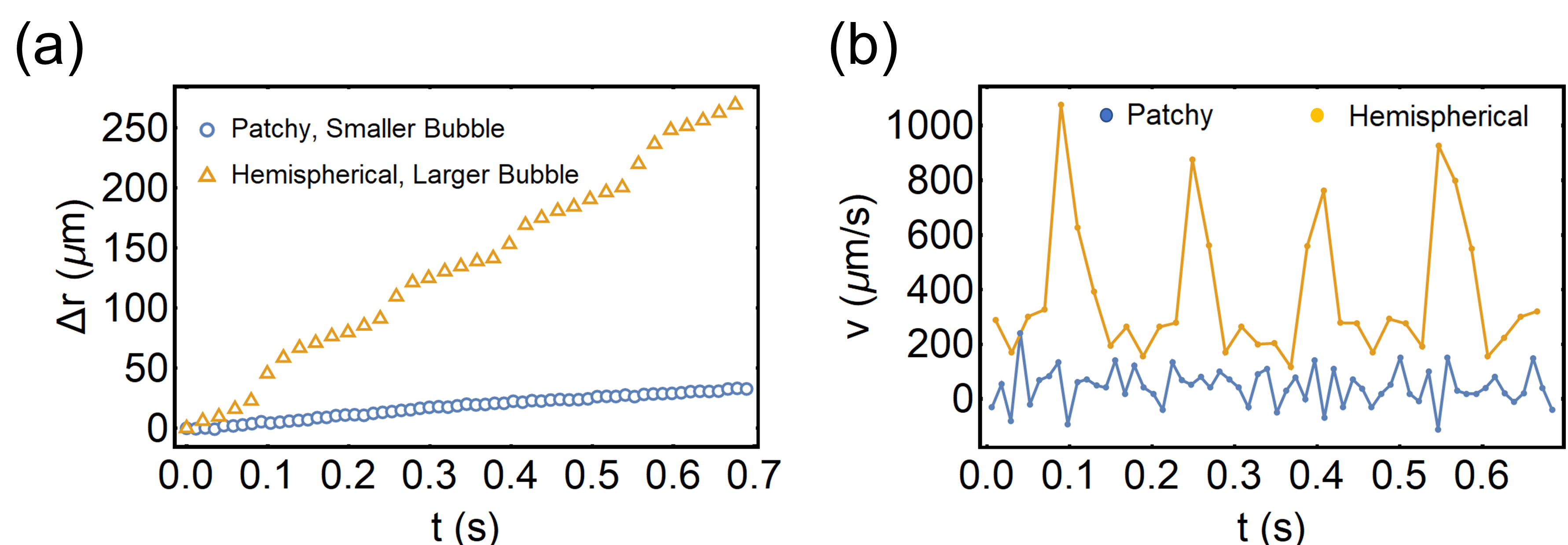}
  \caption {(a) The distance traveled by a hemispherical (yellow) and patchy (blue) micromotor as a function of time. (b) the corresponding instantaneous velocity of the micromotors. For the examples shown, the maximum bubble diameter of the hemispherical micromotor was about 90 $\mu$m, the frequency of bubble production was about 5 Hz, and the concentration of hydrogen peroxide was 30$\%$. The maximum diameter of the bubble produced by the patchy micromotor was about 17 $\mu$m, the frequency was $>$ 19 Hz, and the hydrogen peroxide concentration was 10$\%$.  }
  \label{osPlots}
\end{figure}

\begin{figure}
  \centering
  \includegraphics[width=.95\linewidth]{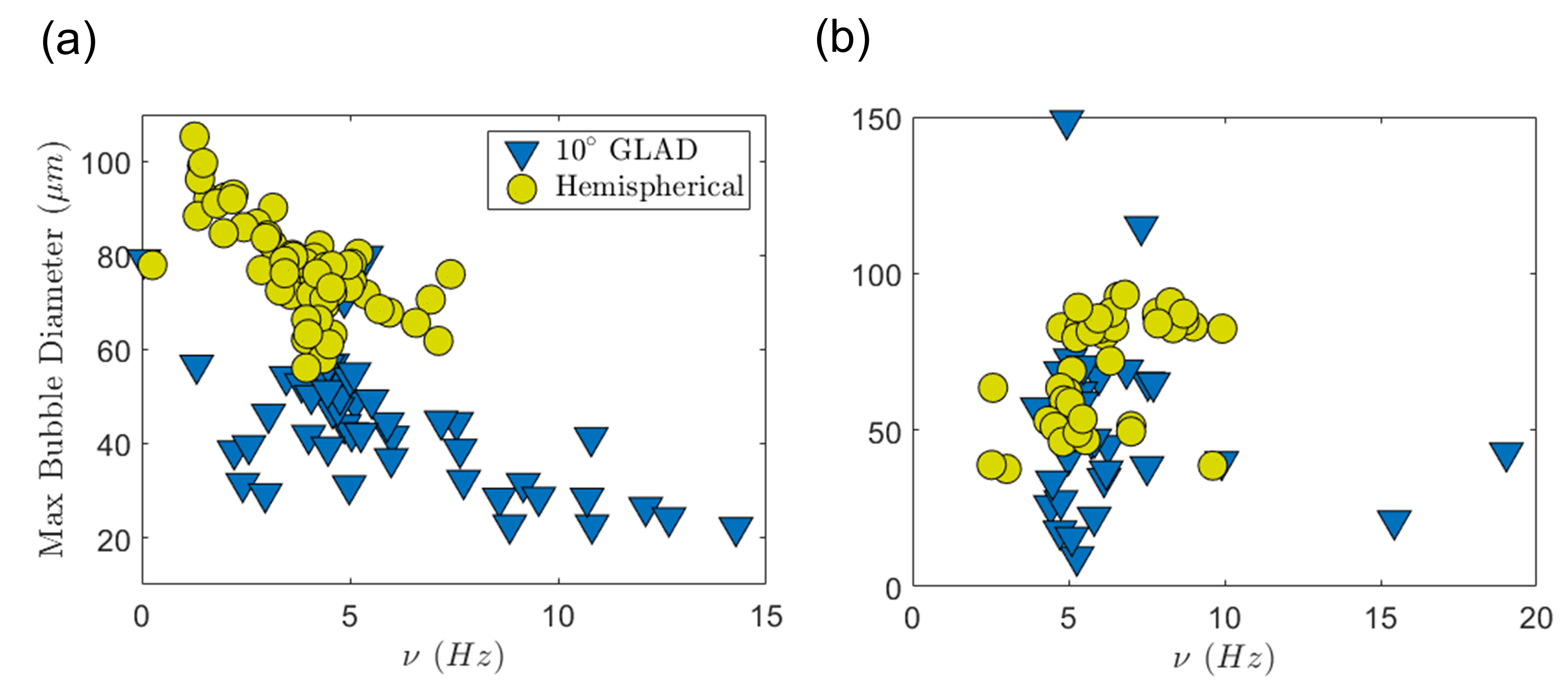}
  \caption {Maximum bubble diameter versus frequency for all (a) temperatures and (b) hydrogen peroxide concentration used. All cases in which the frequency was too fast to measure, which typically only occurred for microrobots with $D_{max} <$ 25 $\mu m$, are omitted.}
  \label{BubbleSzVsFreq}
\end{figure}

The microrobots generally display three types of motion. Microrobots that produce relatively large bubble sizes of roughly $D_{max} > $ 25 $\mu m$ either display exclusively forward motion with a periodic nature or a quasi-oscillatory behavior, as has been described in Refs.~\citenum{ManjarePRL2012,WangLangmuir2018} (see Vid. 1). However, for the microrobots which produce small bubbles of approximately $D_{max} < $ 25 $\mu m$,  this oscillation is not observed and their motion is of a much smoother nature (see Vid. 2). We plot an example of the distance traveled, $\Delta r$, by a patchy microrobot which had a maximum bubble size of about 17 $\mu$m and a frequency of $>$ 19 Hz and a hemispherical microrobot which had a maximum bubble size of about 90 $\mu$m and a frequency of about 5 Hz in Fig.~\ref{osPlots}(a). The corresponding instantaneous velocity is shown in Fig.~\ref{osPlots}(b). The motion displayed by the hemispherically coated microrobot results in periodic spikes in its measured velocity, whereas the patchy microrobot that generates a smaller bubble displays a much smoother motion. The variations from the mean in the this latter case are likely due to noise since there is no periodicity in the data. We find that this anti-correlation between bubble size and frequency is a general trend (see Fig.~\ref{BubbleSzVsFreq} in the SI), particularly in the cases of $D_{max} <$ 25 $\mu m$ in which the frequency was typically greater than 10 Hz and can surpass the limit of our temporal resolution.
%show plot in SI and mention that data of the really small bubbles at high freq is omitted since I couldn't measure the freq

The quasi-oscillatory motion that we observe has been explained previously as a consequence of the forces produced on the microrobot by the bubble as it grows and then bursts. Upon bubble burst, a region of low pressure is created which can cause the microrobot to move backwards \cite{ManjarePRL2012}. However, a hydrodynamic jetting flow can also be created in which fluid flow creates a net forward force on the microrobot \cite{WangLangmuir2018}. In the experiments of Ref.~\citenum{WangLangmuir2018} it was also found that the relative amount of backward to forward motion was determined by the maximum bubble size. For bubble radii less than about 0.7 times that of the microrobot, the microrobot moved backwards followed by a moderate forward motion. For somewhat larger size ratios between 0.7 and 1.5, the microrobots either moved backwards, followed by a larger forward motion, or only moved forward. For even larger bubble sizes up to about 1.8 times that of the microrobot, the microrobots only moved backwards. 

\begin{figure}
  \centering
  \includegraphics[width=.95\linewidth]{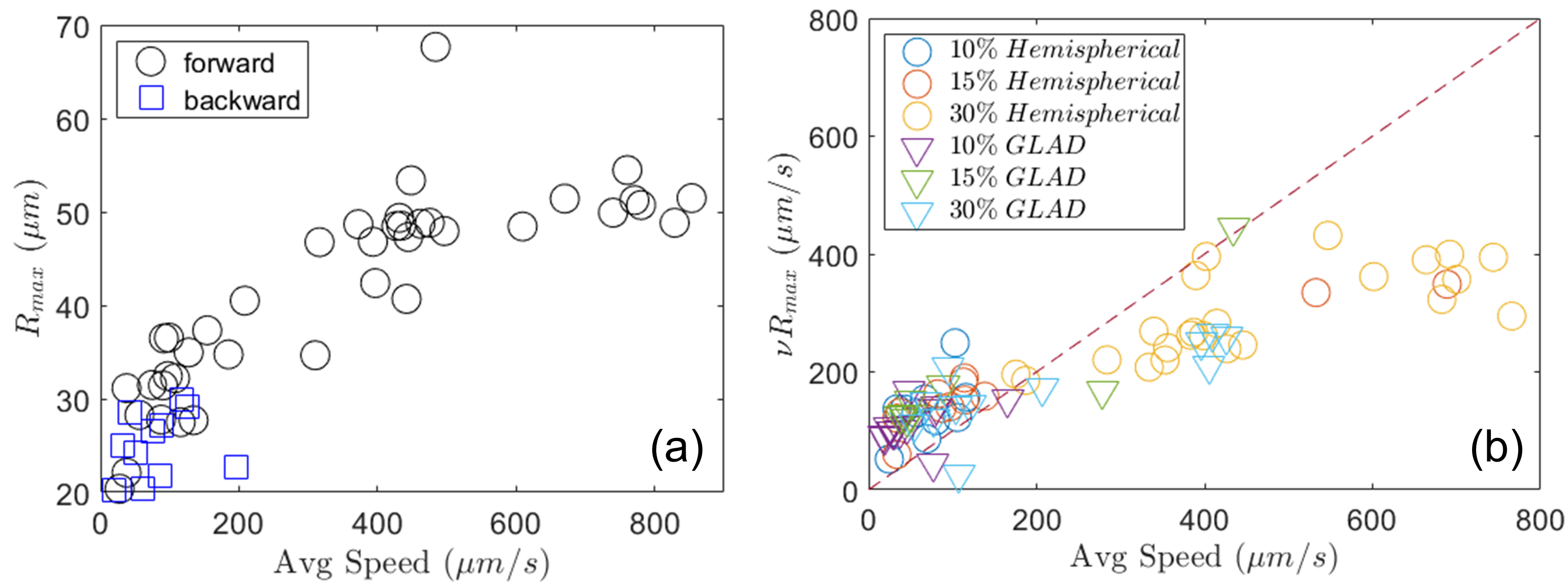}
  \caption {(a) The maximum bubble radius versus average microrobot speed for all cases in which either a forward only (black circles) or backwards (blue squares) motion could be observed following bubble burst. (b) The frequency, $\nu$ of the bubble growth/burst process times the maximum bubble radius, $R_{max}$, plotted against the average microrobot speed. The dashed line has a slope of 1 and corresponds to the case in which the microrobot moves one bubble radius per cycle.}
  \label{freqR}
\end{figure}

To investigate the forward/backward motion and its dependence on bubble size, we examined each video and separated the microrobot dynamics into two categories: those in which the microrobot moved backwards upon bubble burst and those in which it only moved forward. We show a plot of the maximum bubble radius, $R_{max}$, versus average microrobot speed in Fig.~\ref{freqR}(a) for both the GLAD and the hemispherically coated microrobots for cases in which the forward or backward motion could be clearly identified. The blue squares represent cases in which the microrobot moved backward after the bubble burst and the black circles correspond to cases in which the microrobot only moved forward. From the plot, one can see that instances of backwards motion occurred for microrobots that produced bubbles with relatively small radii between 20-30 $\mu m$ and with relatively low velocities of less than about 200 $\mu m/s$. Therefore, our results suggest that for these microrobots, the forward propulsion induced by the hydrodynamic jetting flow has a less significant influence than the backward induced motion created by the pressure decrease upon bubble burst. 

Although we do observe backward motion at bubble to microrobot diameter of a 1.8 ratio, we do not observe the backward motion at the smallest bubble sizes as was observed in Ref.~\citenum{WangLangmuir2018}.  The discrepancy observed could be due to differences in the experimental systems, for example the microrobots used in that study were hollow and hovered beneath the air-liquid interface. Also, the frequency of bubble creation was considerably larger in that study compared to the majority of microrobots we studied. The microrobot size was also somewhat larger (20-50 $\mu m$ diameter compared to our 24 $\mu m$ diameter). We also note that most of our videos were taken at frame rates of 30-50 fps, therefore motion that occurred on timescales less than about 30 ms would be hard to perceive in our videos. 

Aside from the hydrodynamic jetting flow, another propulsion mechanism that drives the microrobot forward is the growth force produced by the bubble as it expands \cite{ManjarePRL2012}. Since the bubbles produced by the microrobots remain attached to the microrobot surface during the growth stage, the growth force can displace the microrobots forward by an amount no greater than the maximum radius of the bubble. Therefore, micro-robots that move with velocities in excess of $\nu R_{max}$, where $\nu$ is the frequency of bubble growth/burst and $R_{max}$ is the maximum radius of the bubble, must be propelled by other mechanisms, such as the hydrodynamic jetting flow. In Fig.~\ref{freqR}(b) we plot the bubble frequency multiplied with the maximum bubble radius versus the average micro-robot speed. The dashed line has a slope equal to one and separates cases in which the microrobot's speed could be attributed solely to a growth force mechanism (on or above the line) from those in which additional mechanisms are required to attain the measured speeds (below the line). As can be seen from the plot, the microrobots with relatively large speeds of more than about 200 $\mu$m/s move faster than what can be attributed to the growth force mechanism alone. The hydrodynamic jetting flow could account for some of the additional propulsion, and indeed the commonality of the approximate speed of 200 $\mu m/s$ at which the forward/backward transition occurred in Figs.~\ref{freqR}(a) seems to further imply that this is the case. We also note, however, that the time over which the jetting flow force persisted in \cite{WangLangmuir2018} was only around 1 ms, whereas we observe forward motion that lasts the entire duration of the bubble growth period (tens of milliseconds) in some cases. This is particularly evident for the fastest micromotors. An example of this can be is shown in Vid. 4 where it can be seen that the center of mass of the bubble itself moves forward during bubble growth. This additional propulsion mechanism following bubble burst is unclear, but could be due to a self-phoretic process, like that responsible for the active motility of smaller, non-bubble propelled, microrobots. If such a process also contributes significantly to the propulsion of microrobots that produce bubbles with $D_{max} < 25 \mu m$, it may account for the relatively large speed of these microrobots compared to those that produce somewhat larger diameter bubbles, as shown in Fig.~\ref{VelVsBubbleSz}(b). %It is interesting to note that the microrobots with $D_{max} < 25 \mu m$ propel at speeds upwards of several hundred micrometers per second despite their small bubble size, suggesting that the mechanism responsible for their propulsion may also not be accounted for by the bubble growth and hydrodynamic jetting flow mechanisms.

Another factor that can affect the resulting microrobot propulsion is the orientation of the microrobot relative to the bubble. In cases in which the bubble/microrobot are oriented along the interface, the resulting propulsion is parallel to the interface and results in a greater velocity. We observe that the relatively slower microrobots appear to reside somewhat below the bubble and therefore a component of the force likely pushes the microrobot downward from the interface, like that observed in Ref.~\citenum{WangLangmuir2018}, which would result in less efficient propulsion along the interface. We note that such a downward propulsion, if present, did not result in the focal plane of the microrobot changing noticeably, however. Also, the sedimentation velocity of the micromotors in the absence of a bubble is such that, for the typical frequencies of bubble production observed of 2-5 Hz, the micromotors would only sink a small fraction of their diameter during a given cycle. Although, fluorescence imaging revealed that the spherical microrobots wobble during each cycle, likely due both to the torque generated on the microrobot as the bubble grows and then bursts and to the rotational torque produced by the weight of the metal cap (see Vid. 3). 

\subsection{Bubble Production Size and Frequency}
\label{bubblesz}

It is relevant to determine how the hydrogen peroxide concentration relates to the reaction rate at the surface of the microrobot and how this might effect the resulting bubble size. The catalytic reaction produces oxygen at the surface of the microrobot. If the concentration of oxygen reaches the saturation level of about 0.28 $mol/m^3$, a bubble is able to nucleate.  Specifically, the rate of oxygen production can be modeled using a  Michaelis-Menten form \cite{HowsePRL2007},

\begin{equation}
    \alpha = \alpha_2\frac{\left[H_2O_2\right]_V}{\left[H_2O_2\right]_V+\frac{\alpha_2}{\alpha_1}}
    \label{MichaelisMenton}
\end{equation}

\noindent where $\alpha$ is the rate of oxygen production, $\alpha_1$ and $\alpha_2$ are catalytic reaction rate constants, and $\left[H_2O_2\right]_V$ is the volume concentration of hydrogen peroxide. The ratio $\frac{\alpha_2}{\alpha_1}$ has been measured to be approximately 0.11 \cite{HowsePRL2007}. Therefore, the reaction rate of hydrogen peroxide at the surface of the microrobot at a hydrogen peroxide concentration of 15$\%$ and 10$\%$ is expected to be about 0.8 and 0.5 of that at 30$\%$, respectively. 

The smaller maximum bubble size in the case of lower hydrogen peroxide concentrations, as well as for the patchy microrobots compared to those with a hemispherical coating as shown in Fig.~\ref{bubblehisto}, could be due to several factors. Previous studies have conjectured that the maximum bubble size is generally determined by the bubble radius at which the amount of oxygen leaving the bubble through diffusion balances the amount entering the bubble via the catalytic reaction at the microrobot's surface \cite{ManjarePRL2012,WangLangmuir2018}. The amount of oxygen that is lost by the bubble increases as the surface area of the bubble increases, therefore there is a threshold maximal size at which the flux of oxygen out of the bubble becomes larger than that into the bubble and the bubble becomes unstable and can collapse. Oxygen that diffuses away from the bubble/microrobot also undergoes transport due to fluid flow produced by the moving microrobot and bubble growth/burst, making for a generally complex process. However, this qualitative model could nevertheless explain the decrease in bubble size at lower hydrogen peroxide concentrations. Also, since the patchy microrobots have about 1/4 the amount of platinum on their surface, the oxygen production rate is presumably also lower which could be responsible in part to the smaller bubble sizes. In addition, the small patch of platinum leads to a more localized region of oxygen concentration and hence presumably a greater gradient of oxygen concentration around the microrobot. This could lead to a smaller region in which there is a sufficient amount of oxygen to sustain bubble growth. 

\begin{figure}
  \centering
  \includegraphics[width=.75\linewidth]{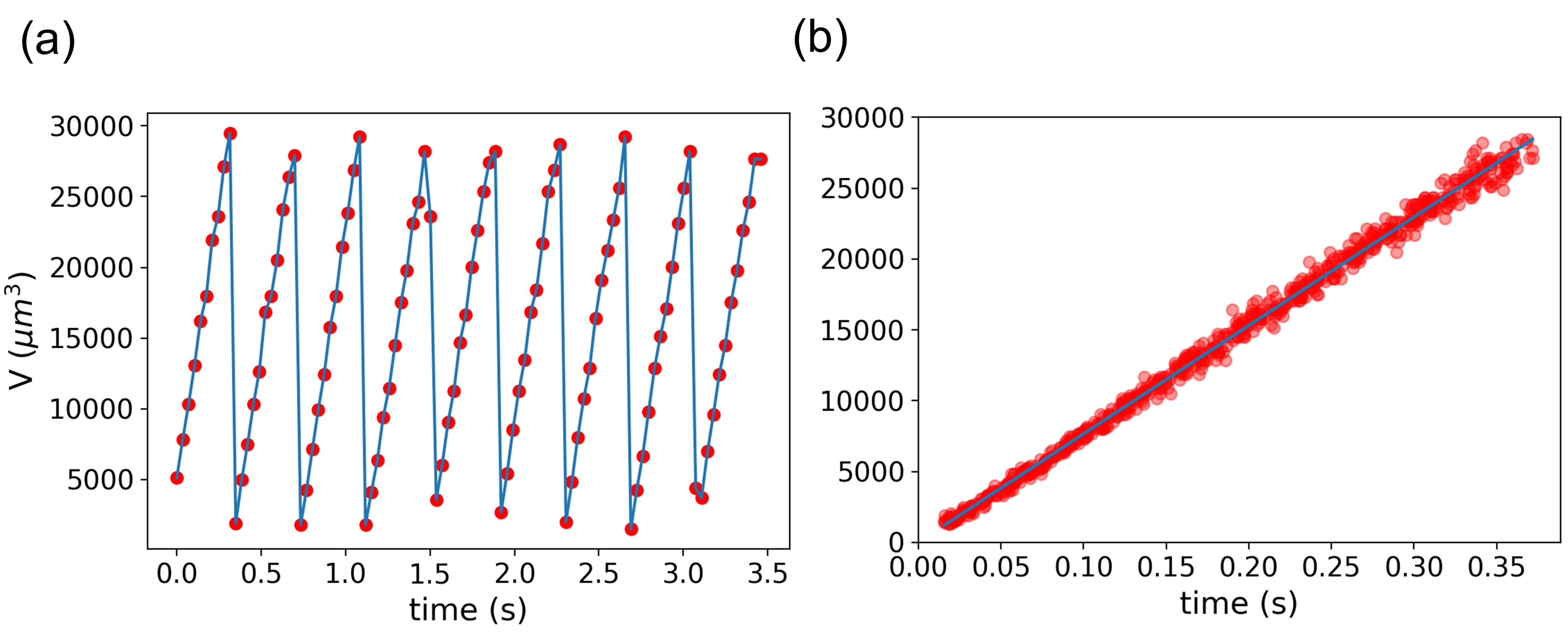}
  \caption {(a) An example of the estimated bubble volume versus time for a microrobot in 30$\%$ $H_2O_2$ that produced a bubble of maximal diameter of around 39 $\mu$m at a frequency of 2.6 Hz. (b) Data from multiple bubble growth/burst periods collapsed onto one curve. Also shown is a linear fit to the data.}
  \label{BubbleSzVsTime}
\end{figure}

In order to further asses this theory, we estimate the volume of the bubble as a function of time which is proportional to the oxygen within the bubble. We determine the volume by measuring the 2D-image of the bubble in each frame of the video and assume a spherical shape of the bubble. Note that, for the diameters considered here, the pressure is constant and is equal to the atmospheric pressure to a good approximation \cite{WangLangmuir2018}. Therefore, the volume of the bubble is proportional to the amount of oxygen inside the bubble via the ideal gas law, $PV = Nk_BT$, where $P$ is the pressure, $V$ is the volume, $N$ is the number of oxygen molecules, $k_B$ is Boltzmann's constant, and $T$ is the temperature. We plot an example of one of these measurements in Fig.~\ref{BubbleSzVsTime}(a). Because of the relatively few number of data points in a given bubble growth process, we combine the data from multiple growth cycles in order to improve the temporal resolution of the data. This was done by conducting a linear fit to the data for each bubble growth cycle and shifting the data along the time-axis such that the intercepts of the linear fits passed through the origin. We show this result in Fig.~\ref{BubbleSzVsTime}(b) along with a linear fit to the data. As one can see from the plot, the bubble volume is linear with time, indicating that the rate of oxygen entering the bubble is approximately constant over the lifetime of the bubble. Since there is no indication of a flux balance prior to bubble burst, which would be indicated by a flattening of the data, other factors may be involved in initiating the bubble burst in our experiments, at least in cases of $D_{max} > 25 \mu m$. 

We note that the bubble size is too small and the frequency too large to obtain accurate data of the bubble size versus time in the case of the subset of microrobots with bubble sizes of $D_{max}< 25 \mu m$, therefore it is possible that the maximal bubble size is determined by an oxygen flux-balance in those cases. One can show that the pressure on the bubble due to surface tension, $2\sigma/R$ (where $\sigma = 72 mN/m$ is the surface tension and R is the bubble radius), starts to become important for bubble diameters of this small size \cite{WangLangmuir2018}. The surface tension is equal to atmospheric pressure of $10^{5}$ Pa at a bubble diameter of about 3 $\mu m$. Therefore, surface tension would effect the growth of these small bubbles and would have to be taken into account.

It also is interesting to note that the volume of a stationary bubble in a liquid that is uniformly supersaturated with oxygen is expected to grow non-linearly with time \cite{Duda1969,LhuissierJFM2012}. Therefore, the measured linearity of the volume of the bubble with time implies, perhaps not surprisingly given the expected non-uniform distribution of oxygen around the microrobot, that this is a poor model for the process by which oxygen enters the bubble. Instead, the linearity suggests that, assuming the rate of oxygen production is constant, the rate of oxygen absorption by the bubble is proportional to the amount of oxygen produced at the microrobot's surface. %I hope this is right

An additional possibility is that the bubble burst occurs due to contact with the air at the interface. Such a process of bubble bursting is described in Ref.~\citenum{LhuissierJFM2012}, for example. The bubble bursting in this case is theorized to be a stochastic process. The bubble size could then be explained by the growth rate of the bubble, i.e. a slower growth rate would lead to a more probable burst occurring before the bubble has time to grow to a larger size. This also would help explain another observation we made with hemispherically coated hollow silica spheres of diameters of 45-85 $\mu m$. We found that these microrobots produced small bubbles and moved at very fast speeds, upwards of around 1 mm/s (see Vid. 5). According to the balance of oxygen flux model, one might expect that the larger surface coverage would result in the production of larger bubbles, but the lack of this observation may indicate that a different process is needed to explain the small bubble sizes. %a similar process is at work in producing small bubbles in the case of the 45-85 $\mu m$ microrobots as that responsible for the small bubble production in the case of the 24 $\mu m$ microrobots.
%[You'd expect that the ones with the largest bubbles would be the highest, and hence burst more often also, but idk. The characteristic timescale in that paper goes as sqrt of R, so larger bubble live longer actually.]
%note that that paper also says that if R is much less than the capilary length, sqrt(sigms/(rho*g)), then the bubble will remain spherical as it contacts the interface. This is about 2 mm so ours would be spherical.

However, the consistency of the maximum bubble size as shown in Fig.~\ref{BubbleSzVsTime} implies that a stochastic process is unlikely to be responsible for the bubble bursting mechanism. We conjecture that several other factors could be responsible for initiating bubble burst, such as detachment of the bubble from the microrobot caused either by bubble buoyancy or growth force which could cause the bubble to overcome the surface adhesive force and move away from and detach from the microrobot. This in turn could lead to bubble instability and burst. These are posited in Ref.~\citenum{Zeng1993}, for example, to explain the size of bubbles in a boiling system.  %Other possible mechanisms that could result in the bubble detaching from the microrobot include fluid flow created by the oxygen production at the surface or the growth of other bubbles which could act to push the larger bubble off of the microrobot's surface. %these last two are pure guesses on my part and might not be reasonable at all

\subsection{Temperature Dependence}
\begin{figure}
  \centering
  \includegraphics[width=.55\linewidth]{ 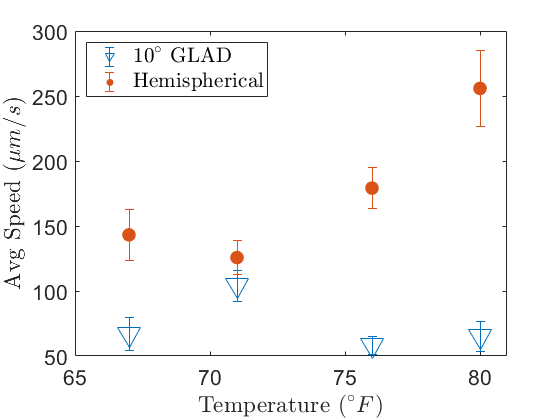}
  \caption The average velocity of the microrobots as a function of temperature. All experiments were conducted at a hydrogen peroxide concentration of 30$\%$. 
  \label{VelVsTemp}
\end{figure}

We studied the temperature dependence of the microrobot dynamics at temperatures of 67, 71, 76, and 80 degrees Fahrenheit at 30$\%$ hydrogen peroxide concentration. The average velocity of the microrobots at each temperature is shown in Fig.~\ref{VelVsTemp}. The average velocity of the patchy microrobots shows little dependence on temperature over the temperature range studied. The hemispherically coated microrobots increase their speed by about a factor of two from 71 ${^\circ}$F to 80 ${^\circ}$F, however. 

\begin{figure}
  \centering
  \includegraphics[width=.95\linewidth]{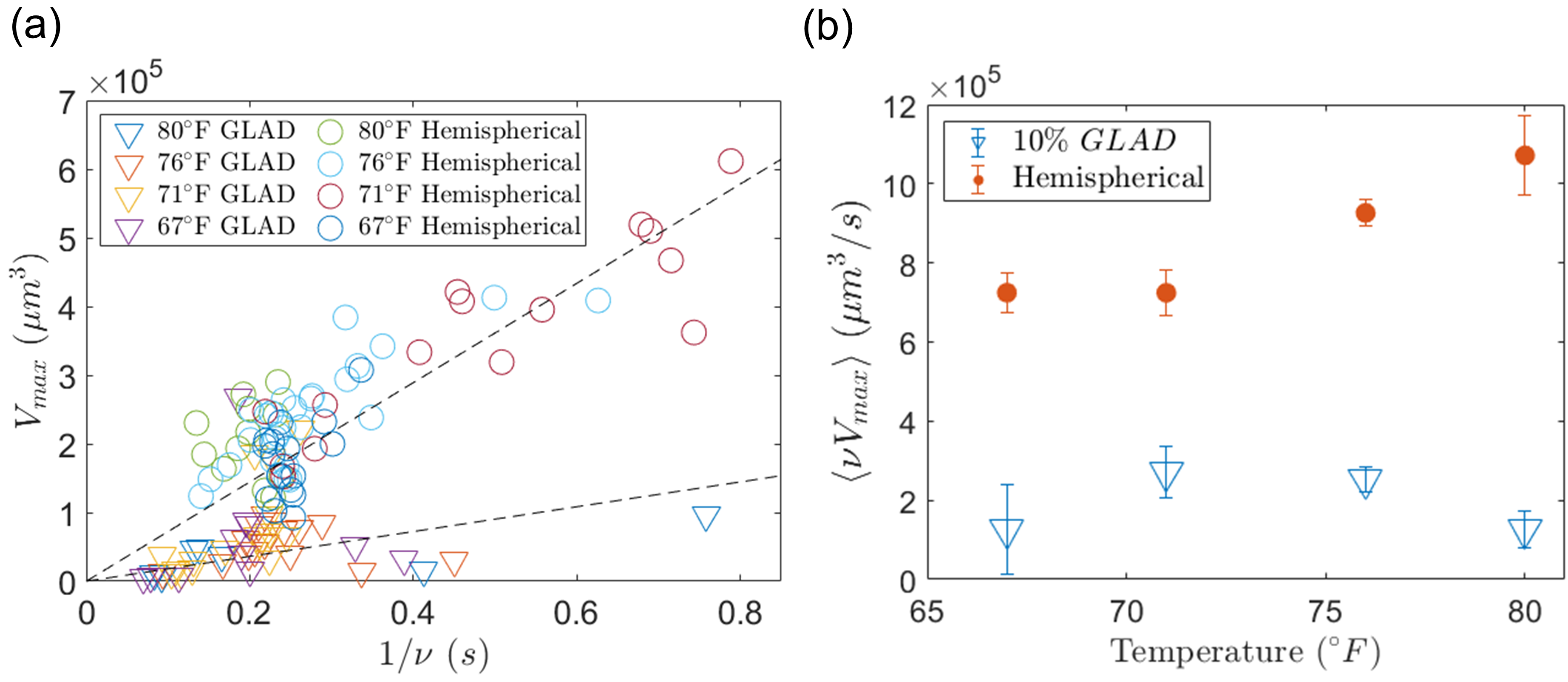}
  \caption (a) The estimated maximum volume of the bubble produced by a microrobot versus the inverse of the bubble frequency for the different temperatures used in the experiments. The dashed lines are a guide to the eye and represent constant volume expansion rates. The slope of the lines differ by a factor of four, equal to the fractional difference in platinum surface coverage of the hemispherical to patchy microrobots. (b) The ensemble median estimated maximum bubble volume times the bubble frequency, which we expect to be proportional to the rate of oxygen produced at the surface of the microrobot. The median rather than the mean was used to avoid outliers effecting the results. Error bars represent standard error of the mean.
  \label{OxygenRate}
\end{figure}

To determine how the oxygen production rates depend on temperature and on the amount of platinum surface coverage of the patchy and hemispherically coated microrobots, we estimate the oxygen production rate produced at the microrobot’s surface assuming that it is proportional to the rate of oxygen increase within the bubble. Because of the linear increase in bubble volume with time, as shown in Fig.~\ref{BubbleSzVsTime}, the volume expansion rate of the bubble can simply be found from $\nu V_{max}$, where $V_{max}$ is the maximum bubble volume and $\nu$ is the frequency of bubble growth. We plot $V_{max}$ versus $1/\nu$ in Fig.~\ref{OxygenRate}(a) for all microrobots at each temperature, and also the median of $\nu V_{max}$ in Fig.~\ref{OxygenRate}(b). Interestingly, the nearly linear relation between $V_{max}$ and $1/\nu$ in Fig.~\ref{OxygenRate}(a) indicates that despite the differences in maximum bubble sizes and frequencies observed in the experiments, the rate of bubble growth was fairly consistent at any given temperature. This indicates, for example, that a smaller bubble size correlates with a smaller bubble lifetime but not with a smaller or larger oxygen production rate. The two dashed lines shown in  Fig.~\ref{OxygenRate}(a) are guides to the eye and represent constant bubble volume expansion rates. The slopes of the two lines differ by a factor of 4, which is approximately equal to the ratio of surface coverage of the hemispherically coated to patchy microrobots. One can see from Fig.~\ref{OxygenRate}(b) that the ratios of $\langle \nu V_{max} \rangle$ for the hemispherically coated to patchy microrbots are similar to this factor across all the temperatures studied. Additionally, the bubble volume expansion rate of the hemispherically coated microrobots increases by about a factor 1.5 from the lowest to the highest temperatures. This is a similar factor to that expected from the temperature dependence of reaction rates which generally increase by a factor of about 1.4 for a change in temperature of 10 ${\circ}$ F.

\begin{figure}
  \centering
  \includegraphics[width=.95\linewidth]{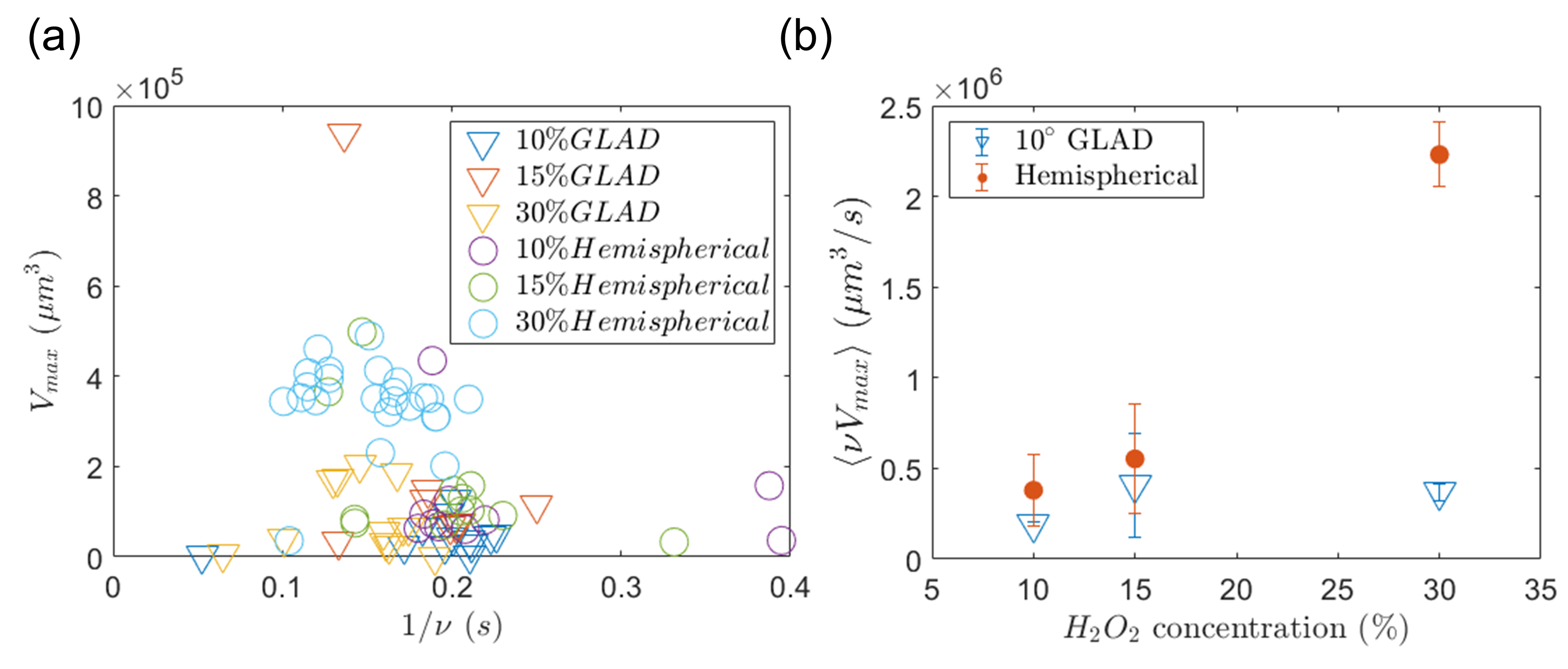}
  \caption (a) The estimated maximum volume of the bubble produced by a microrobot versus the inverse of the bubble frequency at the different hydrogen peroxide concentrations used in the experiments. (b) The ensemble median estimated maximum bubble volume times the bubble frequency, which we expect to be proportional to the rate of oxygen produced at the surface of the microrobot. The median rather than the mean was used to avoid outliers effecting the results. Error bars represent standard error of the mean.
  \label{OxygenRateH2O2}
\end{figure}

Similar plots for different hydrogen peroxide concentrations are plotted in SI Fig.~\ref{OxygenRateH2O2} where we find that the ratio of bubble volume expansion rates in the case of the hemispherically to patchy coated microrobots is smaller than a factor of 4 at the two lowest hydrogen peroxide concentrations of 10 and 15 $\%$. Unexpectedly, the patchy microrobots display little to no change in their volume expansion rate as a function of hydrogen peroxide concentration or temperature. This could be due to differences in the effectiveness of the platinum catalyst at different hydrogen peroxide concentrations for these two types of microrobots. The distribution and thickness of the platinum on the patchy microrobots is less uniform than that on the hemispherically coated microrobots, which could result in differences in effectiveness of the platinum catalyst. Another possibility is that the ratio of oxygen content within the bubble to that produced at the microrobot's surface varies with the concentration of hydrogen peroxide, and that this variation is different for the patchy and hemispherically coated microrobots. %For example, it was found that the microrobots required more time to reach a steady-state of bubble production at lower concentrations of hydrogen peroxide, indicating that their effectiveness was hindered at the lower concentrations. 
%For example, at low concentrations of hydrogen peroxide, the rate of oxygen production is lower and this may allow more time for oxygen to diffuse into the liquid medium rather than entering the bubble. [I think this would result in the opposite effect of what we see in the plot] I think we need a better explanation for this. 

% Things to still talk about:
% turning some bots "on" with other bots
%what about estimating diffusion of o2 out of bubble using standard diffusion equation
%any interest in mentioning a peclet number, diffusion to advection?

%However, the surface area of the $10^\circ$ GLAD patchy microrobots that are covered in platinum is about 1/4 of that of the hemispherically coated microrobots \cite{PawarLangmuir2008}. Therefore, if only the reduced oxygen production rate was responsible for the difference in bubble size, one would expect similar results for the hemispherically coated microrobots at very low hydrogen peroxide concentrations. 
%Instead, we observe that the velocity of the hemispherical microrobots at 10$\%$ $H_2O_2$ concentration is substantially lower than that of the patchy microrobots at 30$\%$ $H_2O_2$ concentration (see Fig.~\ref{speedvsconc}), and that there is little difference in speed between the patchy and hemispherically coated microrobots at the two lowest hydrogen peroxide concentrations. We also find that a relatively large percentage of the patchy microrobots create bubbles with sizes less than the microrobot diameter of 24$\mu$m at 30$\%$ $H_2O_2$ concentration compared to the lower concentrations (see Fig.~\ref{bubblehisto}, which is the opposite of what one might expect if the bubble size was determined solely by reaction rate.

%\subsection{Other Glancing Angles}

\subsection{Interfacial Gravitaxis}
We find that there is a tendency for the micromotors to orient such that they point towards the edges of the droplet. We attribute this effect to the buoyancy of the bubble creating a rotational torque on the microrobot which orients it perpendicularly to the gradient of the slope of the interface. This favored orientation results in a net propagation of the microrobots down the interface and towards the edges of the droplet, thereby demonstrating behavior analogous to positive gravitaxis of  microrobots which orient with the heavy, coated, side facing downwards \cite{CampbellLangmuir2013}. The effect is even more pronounced nearer the edges of the droplet, likely due to an increase in surface curvature there. The microrobots gather at the edge where they maintain a quasi-stationary position, presumably due to a balance of buoyant and propulsive forces. An example of this interfacial gravitaxis behavior is shown in Vid. 8 which displays a micromotor that is magnetically rotated such that it propagates away from the droplet edge and then is allowed to orient back into its preferred state facing toward the edge of the droplet upon zeroing the magnetic field. 

\subsection{Addition of Surfactant}

The addition of surfactant to the solution resulted in a higher frequency of bubble production and a decrease in the size of the bubbles created by the microrobot, in accordance with previous reports \cite{WredeSmall2021}. However, it also led to the accumulation of bubbles in the solution, thus rendering this method ill-suited for micro-manipulation applications where the bubbles would impede the process. A video showing the micromotor in a solution with surfactant is shown in Vid. 9.

\section{Discussion}
We have studied the dynamics of both hemispherically coated and GLAD coated "patchy" spherical bubble-propelled microrobots at an air-liquid interface at various temperatures and hydrogen peroxide concentrations which revealed intriguing and novel behavior of both types of microrobots. We found that a subset of the microrobots produce bubbles of very small size and that this phenomenon is enhanced in the case of the patchy microrobots. Such small bubble sizes are useful for micromanipulation, for example. We have demonstrated directed microassembly and manipulation of passive objects both on a substrate and at the air-liquid interface using precisely controlled patchy bubble-propelled spherical micromotors. 

Additionally, previously proposed propulsion mechanisms of these microrobots were used to analyze the dynamics of the microrobots. It was found that these generally qualitatively account for microrobot motion, however, we find that in some cases the center-of-mass of the bubble during the growth stage moves forward significantly which has not been adequately explained previously. This other mechanism may be of a self-phoretic origin, but future theoretical work is necessary to explore this further.

We also analyzed the bubble bursting mechanism which revealed unexpected behavior. For example, we find that the onset of bubble burst, except perhaps for the smallest bubble sizes, is not due to an oxygen flux balance as has been proposed previously.  We conjecture that other mechanisms, such as bubble growth force or buoyancy, may cause the bubble to detach from the microrobot leading to instability and burst.

The hemispherically coated microrobots exhibited an increase in speed by about a factor of two between the lower temperatures of 67 and 71 ${^\circ}$F compared to the highest temperature studied of 80 ${^\circ}$F. Their speed also increased by about a factor of 7 between 10 $\%$ and 30 $\%$ hydrogen peroxide concentration. This sensitivity could potentially make these microrobots suitable as sensors. The patchy microrobots respond less strongly to changes in both temperature and hydrogen peroxide concentrations, which could also be useful in cases in which more predictable behavior is desired in settings in which such environmental conditions may be unknown or not well controlled.  

Aside from micro-scale manipulation, another potential use of these magnetically controlled microrobots is in interfacial applications, such as biofilm removal. It has been shown previously that the increased mixing created by bubble-propelled micromotors leads to an increase in biofilm removal efficacy \cite{VillaCellReportsPhysSci2020}. Utilizing the magnetic guidance of these micromotors could potentially also make them useful in removal of biofilms in targeted regions of the films, rather than treating the entire film non-selectively. %but why would you ever want this?

Finally, we observe that the microrobots display a positive gravitaxis towards the edges of the droplet which we attribute to a torque on the microrobot created by the buoyancy of the bubble which causes the microrobot to align and propagate towards lower regions of the droplet. Such an interfacial taxis has not yet been reported to the best of our knowledge. 

We also note that the glancing-angle deposition could more generally be useful for coating multiple different materials on the surface of the microrobots by rotating the substrate between depositions, making them potentially motile in a variety of media or able to perform multiple tasks at once. For example, adding a patch of titanium dioxide or ferrous iron to the surface would allow the microrobots to serve as environmental remediation agents \cite{Wu2017,Kong2018,SolerACSNano2013}. The GLAD technique could also be used to create anisotropic surface coatings on the microrobots. This would presumably lead to rotational torques along with translational forces on the microrobots, and hence produce helical motion of the microrobots. Well defined helical motion has previously been sought after with smaller, diffusiophoretically propelled, microrobots \cite{ArcherSoftMatter2015}, but could be interesting to attain in bubble-propelled systems as well. 

%Studying GLAD microrobots with different deposition angles could lend insight into the bubble propulsion mechanism and will be a subject of our future work. [I'm not sure it would be worth it actually]

\section{Materials and Methods}

\subsection{Microscope and Camera}
An Axioplan 2 upright microscope was used with an Axiovert 503 mono camera for viewing and recording videos of the microrobots. Videos were acquired using a 5x objective (Zeiss) and a frame rate of about 30 fps was generally used, although occasionally higher frame rates were utilized in order to capture greater temporal resolution of the microrobot and bubble dynamics.

\subsection{Experimental Procedures}
Experiments were conducted with glass microscope slides which were cleaned with IPA and Acetone and then plasma cleaned for 1 hour. This ensured that the glass surface was hydrophilic, resulting in the hydrogen peroxide solution spreading evenly on its surface with only a small surface curvature. Having little surface curvature is important in minimizing the component of the buoyant forces on the microrobots tangent to the interface. The vial containing the microrobots was vortexed and then 1.5 $\mu$L was extracted and pipetted into a solution of 30$\%$ hydrogen peroxide. For experiments in which a lower concentration of peroxide was used, the microrobots were first placed in a 30$\%$ solution of hydrogen peroxide for 30-45 minutes prior to diluting and pipetting onto the hydrophilic glass slide. A total volume of 140 $\mu$L of the hydrogen peroxide solution containing microrobots was used in all experiments except those in which the microrobots were used to move passive spheres on the substrate. In this case, only 70 $\mu$L total solution was used and time was allowed for some of the liquid to evaporate before conducting experiments. Based on the volume of liquid, we estimate that the thickness of the liquid before evaporation is approximately 200 $\mu m$ and 100 $\mu m$ when using 140 $\mu$L and 70 $\mu$L of the solution, respectively. The glass slide was cut into a 1 inch by 1 inch square and placed under the microscope onto a custom-built stage that contains an array of four home-built electromagnets, arranged along the x and y axes. 

Manipulation of passive spheres at the air-liquid interface was conducted by adding hollow glass microspheres (0.21 g/cc density) with a diameter of 45-85 $\mu m$ (Cospheric) to the hydrogen peroxide solution containing microrobots.

\subsection{Microrobot Behavior Prior to Steady-State}

Sometimes time was required for the microrobots to reach their maximal bubble production rate after adding them into the peroxide solution and observing them on the slide. Therefore, videos were acquired once the microrobots reached a steady-state in which they displayed consistent behavior and approximately maximal speed. We observed that the microrobots began generating bubbles soon after adding them to the substrate at 30$\%$ hydrogen peroxide concentration but took longer to reach a steady-state at lower concentrations. The time required was generally approximately 30 minutes, although occasionally a longer waiting time was needed at the lower hydrogen peroxide concentrations.  

We note that occasionally some micromotors produced large bubbles that continued to grow without bursting for an unusually extended period of time, on the order of a minute or so. This is particularly prevalent in the first several minutes of adding the micromotors to the glass slide before reaching steady-state. Due to buoyant forces, these microrobots aggregated near the center of the slide where the interface is at its highest point. 

\subsection{Electromagnetic Control System}
In the absence of an applied magnetic field, the micromotors tended to undergo circular trajectories, although some microrobots demonstrated motion both to the right and to the left relative to their orientation, therefore performing trajectories resembling a persistent random walk. We applied magnetic fields to control the directionality of the microrobots.

The magnetic system was custom made and is capable of generating magnetic fields along any direction in the x-y plane with fields strengths at the sample’s center up to approximately 10 mT, although generally values of around 5 mT were used when guiding the microrobots. Custom made Matlab code was used to control the strength and direction of the magnetic fields generated by the electromagnets. The code was integrated with an Xbox controller, and a joystick on the controller was used to adjust the magnetic field strength and direction quickly and easily, providing a responsive system that allows for directing the microrobots while observing them in real-time. 

\subsection{Temperature Control}
The temperature of the environment was controlled by enclosing and heating the lab space around the microscope. A hair dryer was used to supply heat and the temperature was measured using a digital thermometer near the microscope. Experiments were conducted within 1 degree of the values stated. 

The temperature controlled experiments were conducted approximately 5 months after the hydrogen peroxide controlled experiments, therefore the longer time that the microrobots were immersed in DI water or the differences in environmental conditions, such as humidity, could be responsible for the somewhat lower speed of the microrobots in the temperature controlled experiments compared to those in the hydrogen peroxide controlled experiments at nominally similar temperatures and concentrations of peroxide. We speculate that this is due to a lower efficiency of the catalytic surface of these microrobots, resulting in a slightly reduced oxygen production rate. In the temperature controlled study, the microrobots also required a somewhat longer time to reach maximal velocity.  The qualitative behavior of the microrobots at steady-state in the two experiments was maintained, however. 

\subsection{Video Analysis}

Image analysis of the microrobot’s speed and the bubble size and frequency of growth was performed using custom python code, details of which are provided in the SI. Briefly, image thresholding is performed to locate the microrobot and bubble center of mass and determine the bubble size. The frequency of bubble growth/dissolution was found by computing a fast-Fourier transform of the bubble size as a function of time. 

\subsection{Microrobot Fabrication}
The microrobots were fabricated using 24 $\mu$m diameter fluorescent magnetic polystyrene spheres (Spherotech Cat. No. FCM-20052-2) using an e-beam vapor deposition procedure. First, the spheres were pipetted into a vial and diluted by a factor of 30 in DI water. The vial was then vortexed and 30 $\mu$L of the spheres are pipetted onto the underside of a clean microscope slide. Multiple 30 $\mu$L droplets were placed on the bottom of the slide while maintaining space between each droplet to avoid merging of the droplets. The spheres sedimented and collected at the bottom of the droplets, forming a dense monolayer with only small gaps between spheres. The droplets were then allowed to evaporate which left only the spheres on the glass surface. The spheres were then coated with a 20 nm layer of iron to make them magnetic. This is followed by a 40 nm layer of platinum which was deposited at an angle $\theta$, as depicted in Fig.~\ref{intro}(b). Angles of $\theta = 10^{\circ}$ and $\theta = 90^{\circ}$ were used, resulting in two types of microrobots: those with a patch of platinum on their surface and those with hemispherical surface coverage, respectively.  Two of the micromanipulation videos shown were performed with patchy microrobots that were produced using a glancing angle of 20$^{\circ}$, although these were not used in any of the experiments from which data was acquired. The microrobots were removed from the glass slide by washing the slide with DI water and allowing the run-off to collect in a vial until the vial is filled with approximately 500 $\mu$L of water. 

\section{Supplemental Information}

\subsection{Microrobot Tracking and Analysis}

Tracking of the microrobots as well as determining their bubble size and frequency of growth was performed using custom python code. The code first applies a thresholding which produces a binary image. For brightfield captures, the background was first subtracted from each image. The microrobots were located by finding the center of mass of each connected region. Trajectories were determined after associating this location data with a particular microrobot, which was done using the “link” function in a particle tracking toolkit called “trackpy” \cite{trackpy}. The microrobot speed was found after smoothing the x and y data followed by fitting an interpolation function using scipy’s univariate-spline interpolation algorithm with default settings. The interpolation function was smoothed prior to taking a time-derivative, giving a velocity in the x and y directions. The speed was then found from the square-root of the sum of the x and y velocities squared. Since the center of mass that was found after thresholding generally lies between the center of the bubble and the microrobot, the smoothing of the center of mass data not only reduces noisy artifacts, but also eliminates the propagation of this localization error into the results of the microrobot speed. The bubble size was determined by finding the area of the connected region after thresholding and subtracting out the estimated area of the microrobot in the absence of a bubble. This method was found to be robust, although not very precise for very small bubbles. For bubbles with a maximum size similar to that of the microrobot or smaller, the maximum size was measured by hand. The frequency of the bubble growth for all microrobots was determined by taking a fast-Fourier transform of the bubble size as a function of time and selecting the frequency with the highest amplitude. The maximum bubble size was determined by finding the maximum size in multiple segments of the video, then taking a median of these values. At least 10 microrobots were recorded in at least three different experiments to determine the overall averages shown in the figures.% figures \ref{speedvsconc} and \ref{osPlots}.

\subsection{Additional Figures}

\begin{figure}[!htb]
  \centering
  \includegraphics[width=.6\linewidth]{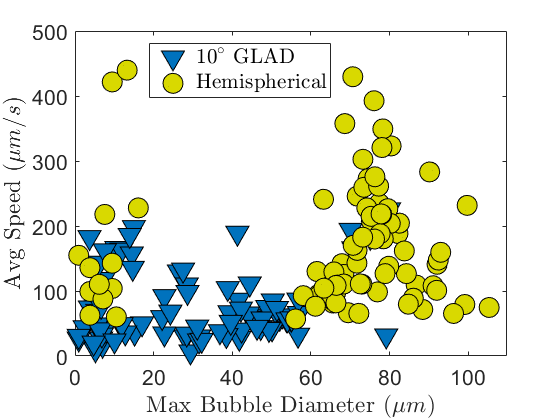}
  \caption {The average speed of the microrobots versus their maximum bubble diameter, for all temperatures studied.}
  \label{VelVsBubbleSzTemp}
\end{figure}

\begin{figure}
  \centering
  \includegraphics[width=.95\linewidth]{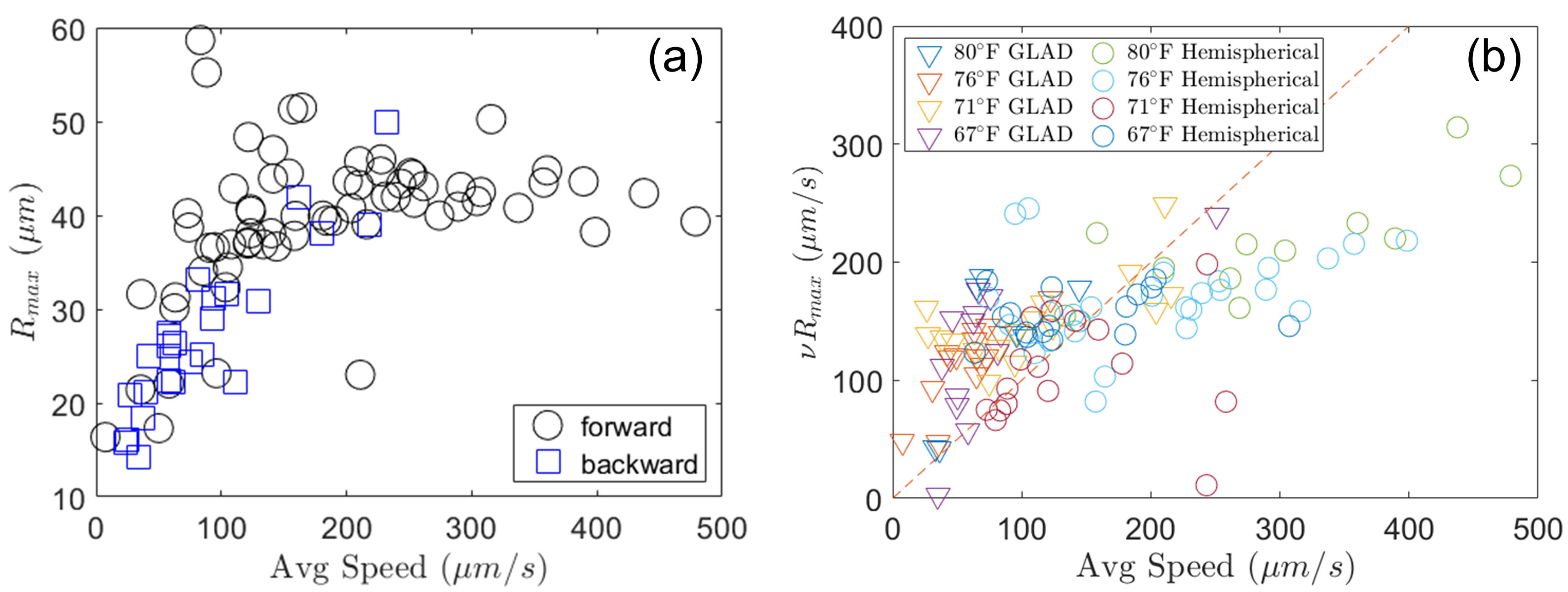}
  \caption {Data shown corresponds to temperature control experiments. (a) The maximum bubble radius versus average microrobot speed for all cases in which either a forward only (black circles) or backwards (blue squares) motion could be observed following bubble burst. (b) The frequency, $\nu$ of the bubble growth/burst process times the maximum bubble radius, $R_{max}$, plotted against the average microrobot speed. The dashed line has a slope of 1 and corresponds to the case in which the microrobot moves one bubble radius per cycle.}
  \label{freqRTemp}
\end{figure}

\begin{figure}
  \centering
  \includegraphics[width=.95\linewidth]{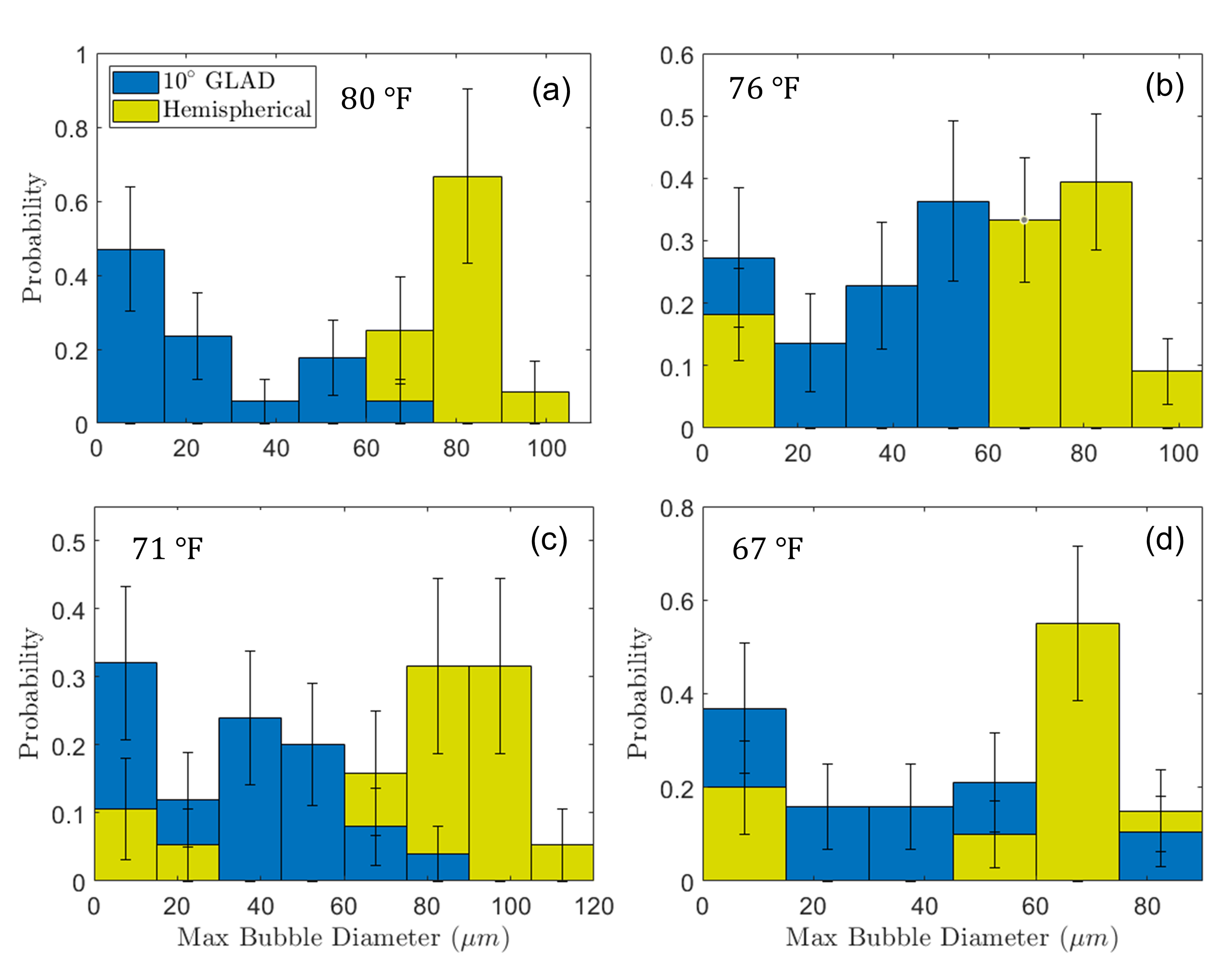}
  \caption {Probability of a micromotor producing a bubble of a given diameter for both the patchy and hemispherically coated microrobots at different temperatures of 80, 76, 71, and 67 ${^\circ}$F in (a-d), respectively. To allow for easier viewing of the data, the larger of the two bars is placed behind the smaller. The error bars correspond to square root of N uncertainties.}
  \label{bubblehistoTemp}
\end{figure}

\begin{figure}
  \centering
  \includegraphics[width=.95\linewidth]{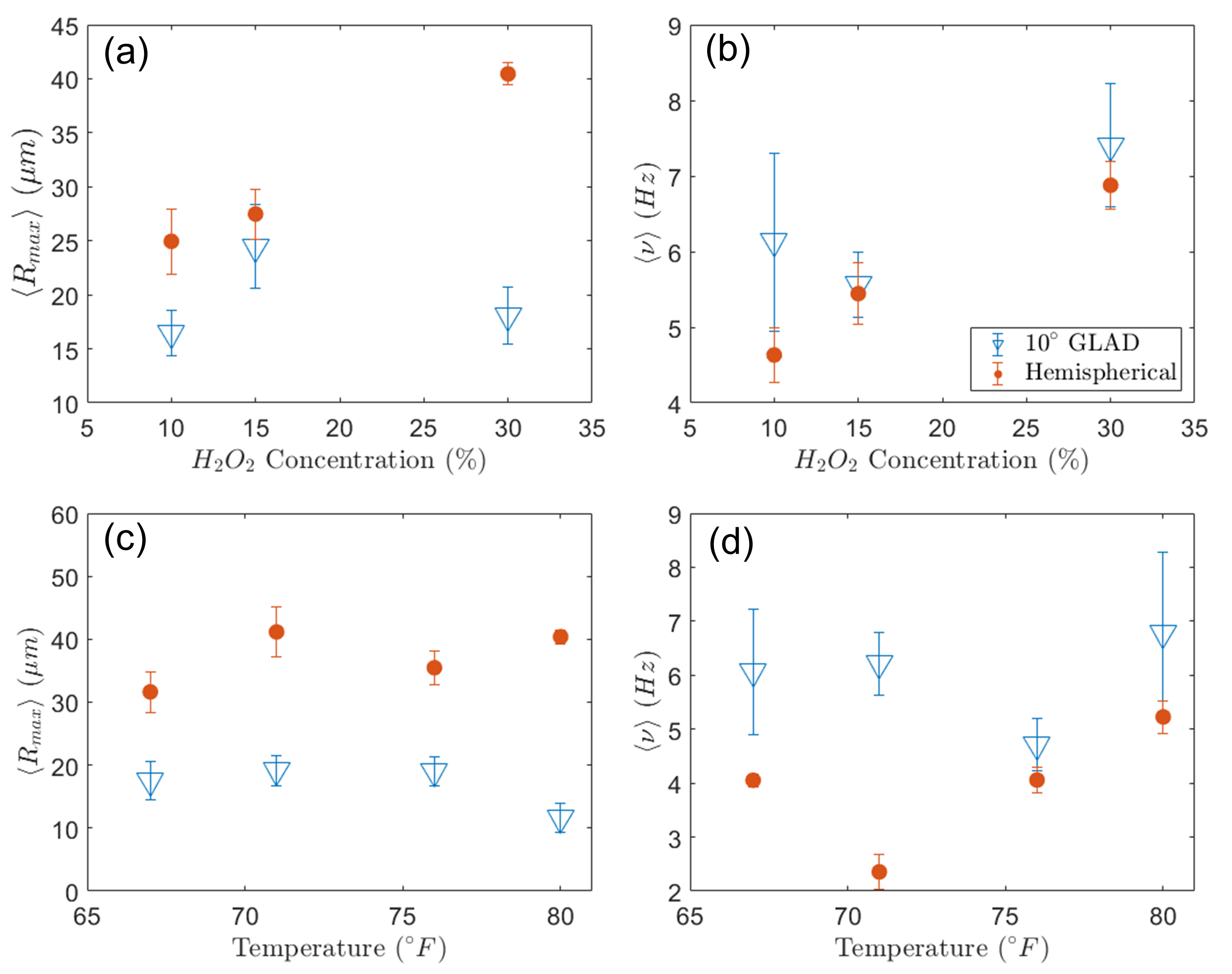}
  \caption {(a,b) Ensemble average maximum bubble radius and frequency, respectively, for all hydrogen peroxide concentrations studied. (c,d) Ensemble averages at each temperature used in the experiments. Note that frequencies that were too high to be measured accurately were omitted from the calculated averages shown.}
  \label{AvgRandFreq}
\end{figure}

\clearpage

\bibliography{scibib}

\bibliographystyle{Science}

% Following is a new environment, {scilastnote}, that's defined in the
% preamble and that allows authors to add a reference at the end of the
% list that's not signaled in the text; such references are used in
% *Science* for acknowledgments of funding, help, etc.

%\begin{scilastnote}
%\item We've included in the template file \texttt{scifile.tex} a new
%environment, \texttt{\{scilastnote\}}, that generates a numbered final
%citation without a corresponding signal in the text.  This environment
%can be used to generate a final numbered reference containing
%acknowledgments, sources of funding, and the like, per {\it Science\/}
%style.  Along those lines, we'd like to thank readers of this document
%for their attention, and invite them to address any questions to
%Stewart Wills, at swills@aaas.org.
%\end{scilastnote}

% For your review copy (i.e., the file you initially send in for
% evaluation), you can use the {figure} environment and the
% \includegraphics command to stream your figures into the text, placing
% all figures at the end.  For the final, revised manuscript for
% acceptance and production, however, PostScript or other graphics
% should not be streamed into your compliled file.  Instead, set
% captions as simple paragraphs (with a \noindent tag), setting them
% off from the rest of the text with a \clearpage as shown  below, and
% submit figures as separate files according to the Art Department's
% instructions.

\clearpage

%\noindent {\bf Fig. 1.} Please do not use figure environments to set
%up your figures in the final (post-peer-review) draft, do not include graphics in your
%source code, and do not cite figures in the text using \LaTeX\
%\verb+\ref+ commands.  Instead, simply refer to the figure numbers in
%the text per {\it Science\/} style, and include the list of captions at
%the end of the document, coded as ordinary paragraphs as shown in the
%\texttt{scifile.tex} template file.  Your actual figure files should
%be submitted separately.

\end{document}